\documentclass[psfig]{aa}

\usepackage{graphicx}
\usepackage{natbib}
\usepackage{subfigure}

\begin{document}

\title{Photometric Estimates of Stellar Masses in High-Redshift Galaxies}

\author{Stefano Berta\inst{1},
Jacopo Fritz\inst{1}, Alberto Franceschini\inst{1}, 
Alessandro Bressan\inst{2} \& Carol Lonsdale\inst{3}
}

\institute{$^1$ Dipartimento di Astronomia, vicolo dell'Osservatorio 2, 35122 
Padova, Italy\\
$^2$ Osservatorio Astronomico di Padova, vicolo dell'Osservatorio 5,
35122 Padova, Italy\\
$^3$ Infrared Processing \& Analysis Center, California Institute of Technology
100-22, Pasadena, CA 91125, USA. 
}

\offprints{Stefano Berta, \email{berta@pd.astro.it}}

\date{Received ..... / Accepted .....}

\titlerunning{Photometric mass estimates in high-z galaxies}
\authorrunning{Berta S., et al. }

\abstract{
We present a new tool for the photometric estimate of stellar masses in
distant galaxies. The observed source spectral energy distributions 
are fitted by combining sets of various single stellar populations, with
different normalizations and different amounts of dust extinction, for a given
(Salpeter) IMF.  
This treatment gives us the best flexibility and robustness when dealing with the
widest variety of physical situation for the target galaxies, including inactive
spheroidal and active starburst systems. 
We tested the code on three classes of sources: complete
samples of dusty ISO-selected starbursts and of K-band
selected ellipticals and S0s in the HDF South, and a representative sample of
$z\sim 2$ to 3 Lyman-break galaxies in the HDF North.  
We pay particular attention in evaluating the uncertainties in the stellar mass
estimate, due to degeneracies in the physical parameters, 
different star formation histories and 
different metallicities. Based on optical-NIR photometric data, the stellar
masses are found to have overall uncertainties of a factor of $\sim 2$ for
E/S0s,  while for the starburst population these rise to factors 2$-$5 (even
including ISO/15$\mu$m photometric data), and up to $\ge 10$ for Ly-break
galaxies. Our analysis reveals in any case the latter to correspond to a galaxy
population significantly less massive ($M<$ a few $10^{10}M_\odot$) than those
observed at lower redshifts (for which typically $M>$ several $10^{10}M_\odot$),
possibly indicating substantial stellar build-up happening at $z\sim 1$ to 2 in
the field galaxy population. 
Using simulated deep SIRTF/IRAC observations of starbursts and Lyman-break
galaxies, we investigate how an extension of the wavelength dynamic range will
decrease the uncertainties in the stellar mass estimate, and find that they will
reduce for both classes to factors of 2$-$3, comparable to what found for E/S0s
and good enough for statistically reliable determinations of the galaxy
evolutionary mass functions.   

}

\maketitle

%%%%%%%%%%%%%%%%%%%%%%%%%%%%%%%%%%%%%%%%%%%%%%%%%%%%%%%%%%%%%%%%%%
% text body

\section{Introduction}

One of the still open critical questions of modern cosmology is to
understand the epoch at which galaxies assembled the bulk of their stellar
content.
In the so-called monolithic scenario, the assembly of galaxies took place on
rapid timescales at high redshifts, then galaxies evolved passively to
present days.
On the contrary, in the hierarchical scenario (Kaufmann \& Charlot, 1998)
galaxy formation is predicted to be a more continuous process and elliptical
galaxies to assemble through merging of lower mass disc galaxies at moderate
redshifts. 

Many authors have analysed the evolution of the global 
star formation rate (SFR) through cosmic history, back to z$\ge$3,
using different observational tracers of the SFR in distant galaxies. It is
generally accepted that galaxies produced stars 
more actively in the past than today, but the true rates of star formation are
affected by a variety of uncertainties and biases, expecially related to the
amount of dust in galaxies and its effect on the SFR tracers.

A complementary approach consists in measuring the dynamical or stellar masses of
distant galaxies, instead of the instantaneous SFR.
Dynamical masses are directly connected to theoretical predictions, but 
very difficult to measure, requiring high spatial
and spectral resolution spectroscopy of selected samples of faint high-redshift
galaxies.
Stellar masses, on the contrary, are less well determined by theory, but have the
advantage that can be derived using multiband optical and near-IR
photometry as a powerful alternative to time-expensive spectroscopic investigations.
Indeed galaxy near-IR SEDs show a moderate dependence 
on the age of the contributing stars (e.g. Franceschini \& Lonsdale, 2003),
because in a typical galaxy the stellar mass is dominated by low-mass stars,
with evolutionary timescales of the order of the Hubble time.  
As discussed by several authors (e.g. Lancon et al. 1999, Origlia \& Oliva 2000),
these moderate-mass stars emit predominantly in the near infrared (NIR, J-to-K
restframe bands), especially during their cool giant phase, 
and are only slightly affected by dust extinction.

In this work we present a new spectro-photometric synthesis tool, aimed at the
estimate of stellar masses in galaxies from a thorough analysis of their spectral energy
distributions (SEDs). The tool combines a set of single stellar populations of
different ages, assuming different star formation rates and different amounts of
extinction for each.
The code is tested by fitting the optical-NIR broadband SEDs
of two samples of intermediate ($z=0.5-1.5$) redshift
galaxies, luminous IR-selected starbursts in the HDFS (Franceschini et al., 2003) and
mostly passively-evolving K-band selected ellipticals from Rodighiero et al. (2001).
A benchmark higher-redshift population, also targeted by our analysis, was
extracted from a sample of $z\sim 2-3$ Lyman-break galaxies (Papovich et al., 2001).

We pay particular attention to the uncertainties in the mass estimates, due to
degeneracies in extinction, age and star formation history, and discuss how
forthcoming near-to-mid infrared data from SIRTF will further constrain the
photometric mass estimate for these sources.

The paper is structured as follows.
Section 2 describes the three samples of selected sources. Section 3 presents the
model tools used to fit the observed optical-NIR spectral energy
distributions of the sources and the different star formation histories assumed
for the two classes of galaxies analyzed. Section 4 discusses our results, with
a particular care for degeneracies. Section 5 shows perspectives for future
SIRTF/IRAC observations and Section 6 summarizes our conclusions.
We assume a $H_0=65$ $[$km s$^{-1}$ Mpc$^{-1}]$, $\Omega_m=0.3$,
$\Omega_\Lambda=0.7$ cosmology.

\section{The samples}

Our analysis compares synthetic spectra of a new spectrophotometric code
with broad-band observational data
to estimate the galaxy's main physical parameters, particularly the 
total stellar mass, but also, at the same time, age, extinction and star-formation rate. 
Our aim here is to investigate in particular the uncertainties in the stellar
mass estimates from degeneracies in the solutions due to the sparse
characterization of the observational spectrum. 

We compare with test cases provided by populations of luminous starburst galaxies
and spheroidal galaxies, the two representing somewhat limiting physical situations.
In the former class, the presence of red giants and dust-reddened young
stars complicates the analysis and widens the uncertainty range. 
On the other hand, for passively evolving spheroids the
uncertainties are expected to be much lower.
%, and basically due to the uncertain stellar IMF.  

A third situation explored here is higher redshift ($z\sim 2-3$) Lyman-break
galaxies, to show how the poorer coverage of the rest-frame spectral energy
distribution is affecting the uncertainties in the parameters.

\subsection{Mid-IR selected starbursts}

The rich variety of high photometric quality data in the Hubble Deep Field South
and its Flanking Fields allows us to build accurate SEDs for galaxies.
We have selected a sample of intermediate--redshift sources detected
by ISO in the LW3 ($\lambda_c=15\ \mu$m) band (Franceschini et al. 2003, Oliver
et al. 2002) to a flux limit of $S_{LW3}=90 \ \mu$Jy. 
Franceschini et al. (2003) compare the ISO source list with
those from the Deep ESO Imaging Survey (EIS Deep, da Costa et al. 1998) and with
the optical UBVRI catalogue by Teplitz et al. (1998), and identify 35
extragalactic objects. 
Of these 35 sources, 3 are likely to be hosting an AGN component and for another 2 not
enough photometric data are available. For the remaining 30 galaxies, UBVRIJHK photometry is
available (Da Costa et al. 1998, Tepliz et al. 1998), as well as the LW3 15
$\mu$m ISO flux and --- in some cases --- the LW2 6.7 $\mu$m flux. For 17
sources spectroscopic redshifts are available from Rigopoulou et al. (2000) and
Franceschini et al. (2003); for the 13 others we rely on the photometric estimate.

\subsection{K-band selected ellipticals and S0s}

As representative of this class we chose a morphologically selected E/S0 sample
by Rodighiero et al. (2001) in the HDFS.  
Based on a SExtractor analysis (Bertin \& Arnouts, 1996) of the EIS-Deep K band SOFI image
(Da Costa et al., 1998), they computed a completeness limit of K$\simeq$20.2
(Vega magnitude). 
Among this list, the subpopulation of spheroidal galaxies was identified 
from both visual morphological inspection and using the authomatic analysis 
tool GASPHOT (Pignatelli \& Fasano, 1999). From this, a sample of 29 E/S0
brighter than K$=$20.1, in the WFPC2 HDFS field was selected. 

Among these sources, data in the whole UV-to-NIR
spectral range are only available for 23 galaxies: UBVI from HST (filters F300W, F450W, F606W, F814W,
Williams et al. 1996) and JHK from EIS Deep (SOFI/NTT, Da Costa et al. 1998)
photometry.
For 10 of these sources spectroscopic redshifts are available from Vanzella et
al. (2003) and Sawicki \& Mallen-Ornelas (2003).
For the remaining, the photometric estimate is reliable to within 5\%.

\begin{figure*}[!ht]
\centering
\includegraphics[width=0.46\textwidth]{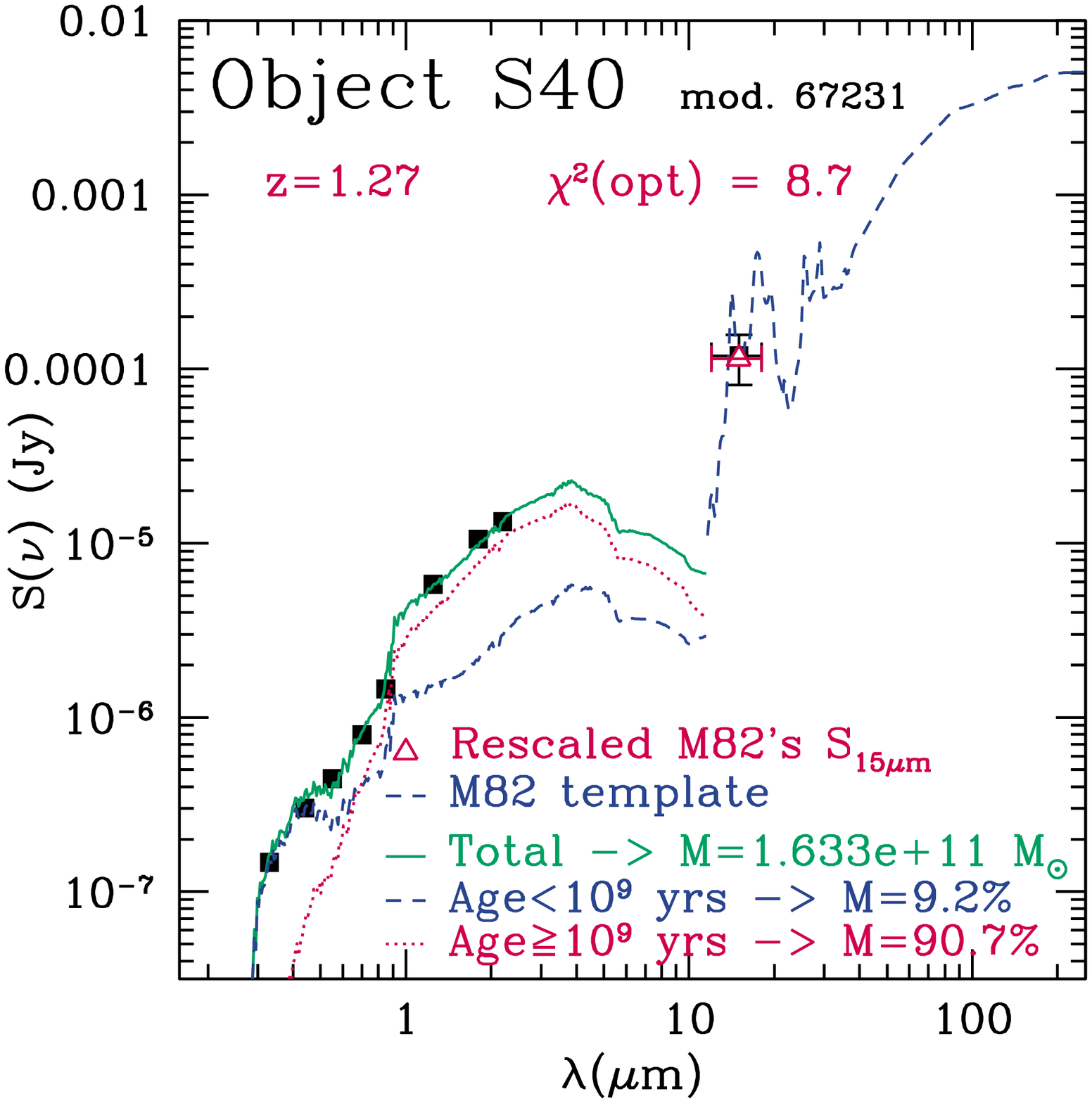}
\includegraphics[width=0.46\textwidth]{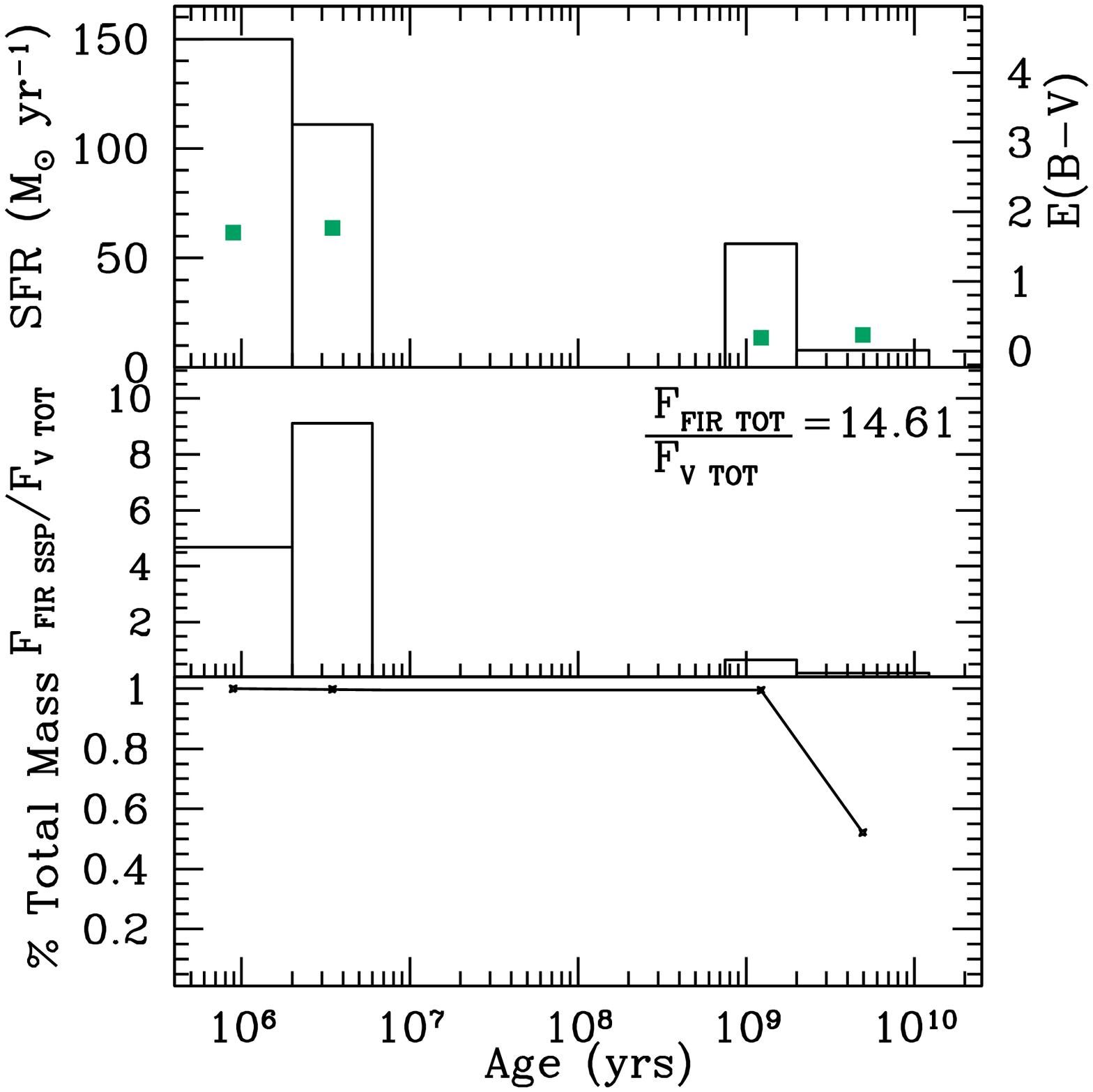}
\caption{HDFS/LW3 source S40 at $z=1.27$: best fit mass is $M=1.63e11$. {\bf Left
Panel}: comparison of the observed SEDs to the best fits. The solid line is the
sum of all the used SSPs; the dashed line
is the sum of all the young SSPs (age $< 10^9$ yrs); the dotted line represents
the old populations (age $\ge 10^9$ yrs); the dashed thick line longward 5
$\mu$m (restframe) is the k--corrected rescaled M82 template; the square black dots
are the observed photometric datapoints; the triangular red point at $\lambda\sim 15 \
\mu$m is the synthetic LW3 flux (overlapping to the observed LW3 when the FIR
constraint is satisfied). {\bf Right Panel} {\em top}: values 
of SFR (histogram, left y--scale) and E(B--V) (dots, right y--scale) for each
single stellar population. {\em Middle}: ratio between
the FIR and V-band fluxes for the different SSPs. This quantity is used to
asses both the contribution of various stellar populations to the FIR luminosity
and the effect of extinction on each population (depressing the optical light
and boosting the IR). The global ratio between the FIR and V fluxes of the
composite final synthetic spectrum is reported on the right. {\em Bottom}:
cumulative stellar mass formed in the galaxy. The youngest stars, significantly
contributing to the FIR bolometric emission, provide almost no contribution to
the total luminous mass assembled in the galaxy.} 
\label{fig:results_s40}
\end{figure*}

\subsection{Lyman break galaxies}

Papovich et al. (2001) report the analysis of stellar populations 
of a sample of Lyman-break galaxies in the Hubble Deep Field North, selected 
through the U dropout tecnique, up to redshift 3.4. 

The publicly available photometric data consist of the original optical WFPC2
F303W, F450W, F606W, F814W HDFN observations by Williams et al. (1996), the
NICMOS F160W and ground based (IRIM) Ks data by Papovich et al. (2001) and 
the J band magnitudes reported by Sawicki \& Yee (1998), obtained with IRIM on
the KPNO 4m Mayall telescope. Redshifts have been
spectroscopically derived by various authors (e.g. Steidel et al. 1996, Lowental
et al. 1997).

We have selected a sub-sample of 8 sources between $z\sim2-3$: this redshift
range is the one where future SIRTF/IRAC observations will be more
effective in constraining the stellar masses of such objects.

\section{The Model Tool}
\label{par:tools}

\begin{figure*}[!ht]
\centering
\includegraphics[width=0.97\textwidth]{dummy_f2.eps}
\caption{{\em Left}: $\chi^2$ contours on the $M$ vs. $L_{FIR}$ diagram, for
source S40. Contours
are drawn at 1,2,3$\sigma$ levels, corresponding to 68.3\%, 95.5\% and 99.7\%
confidence (decreasing as shade darkness increases). The solid and dashed
horizontal lines represent the far--IR luminosity and uncertainty as derived
from the observed LW3 flux, based on fitting 
with the M82 template. {\em Right}: $\chi^2$ values as a function of the stellar
mass of the models, for those solutions reproducing the LW3 flux within 1\%. The
mass uncertanty range is $[0.70,3.80]\ 10^{11}M_\odot$.}  
\label{fig:contours_s40}
\end{figure*}

The observed SEDs for the target galaxies span wide ranges of rest-frame
wavelengths (0.2$-$1.4 $\mu$m at the mean redshift of the starburst sample,
$z=0.6$), and therefore  
are sensitive to stellar populations over a wide range of ages and
mass/luminosity ratios and to different aspects of galaxies star formation
histories. 
In the starburst galaxy case the observed UV-optical
SED is strongly affected by dust extinction. Dust re-radiation is expected in
the  mid- to far-IR. The relative contributions of young red stars from older
stellar populations can then be constrained with ISO mid-IR observations. 

\begin{figure*}[!ht]
\centering
\includegraphics[width=0.325\textwidth]{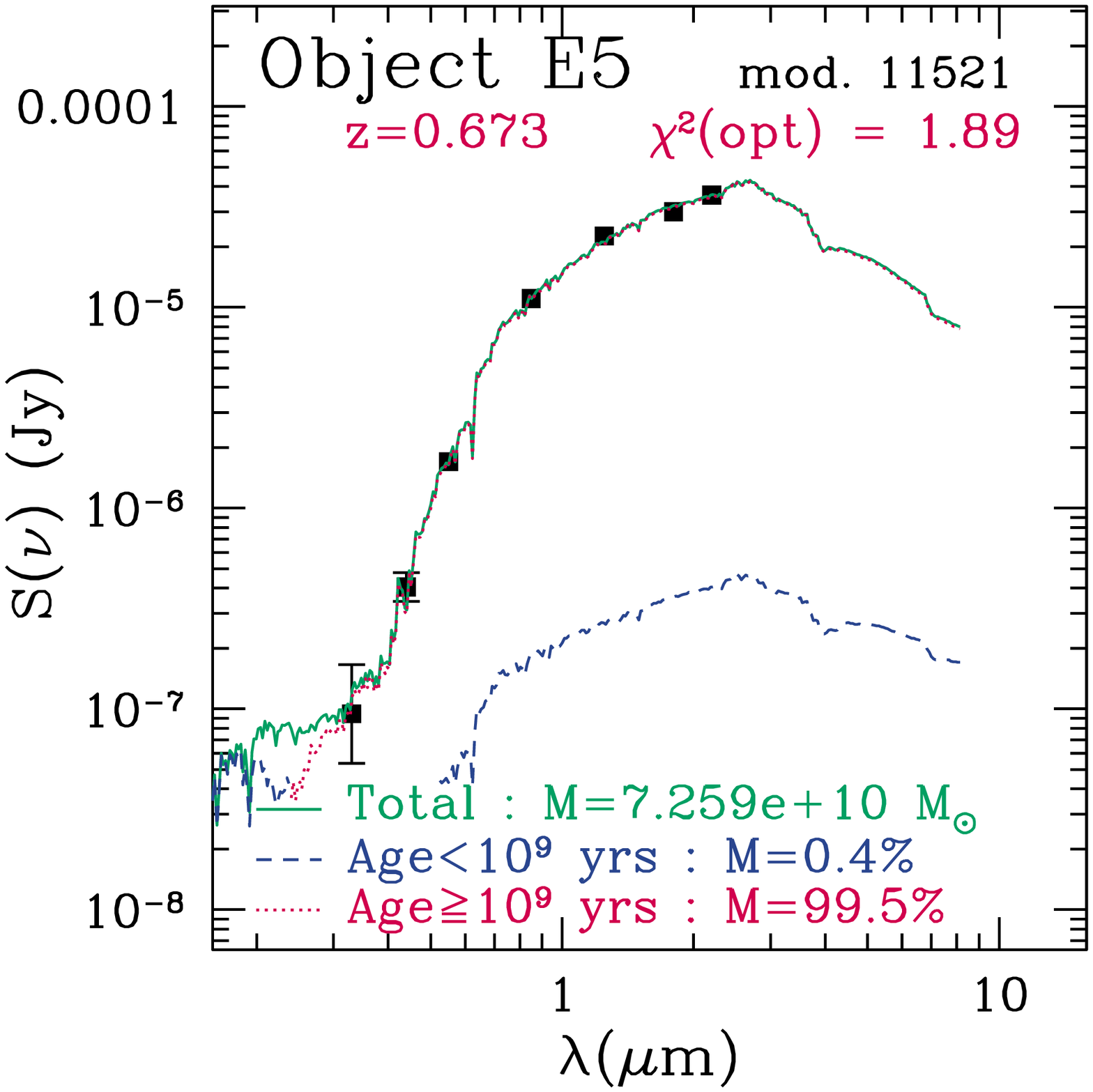}
\includegraphics[width=0.325\textwidth]{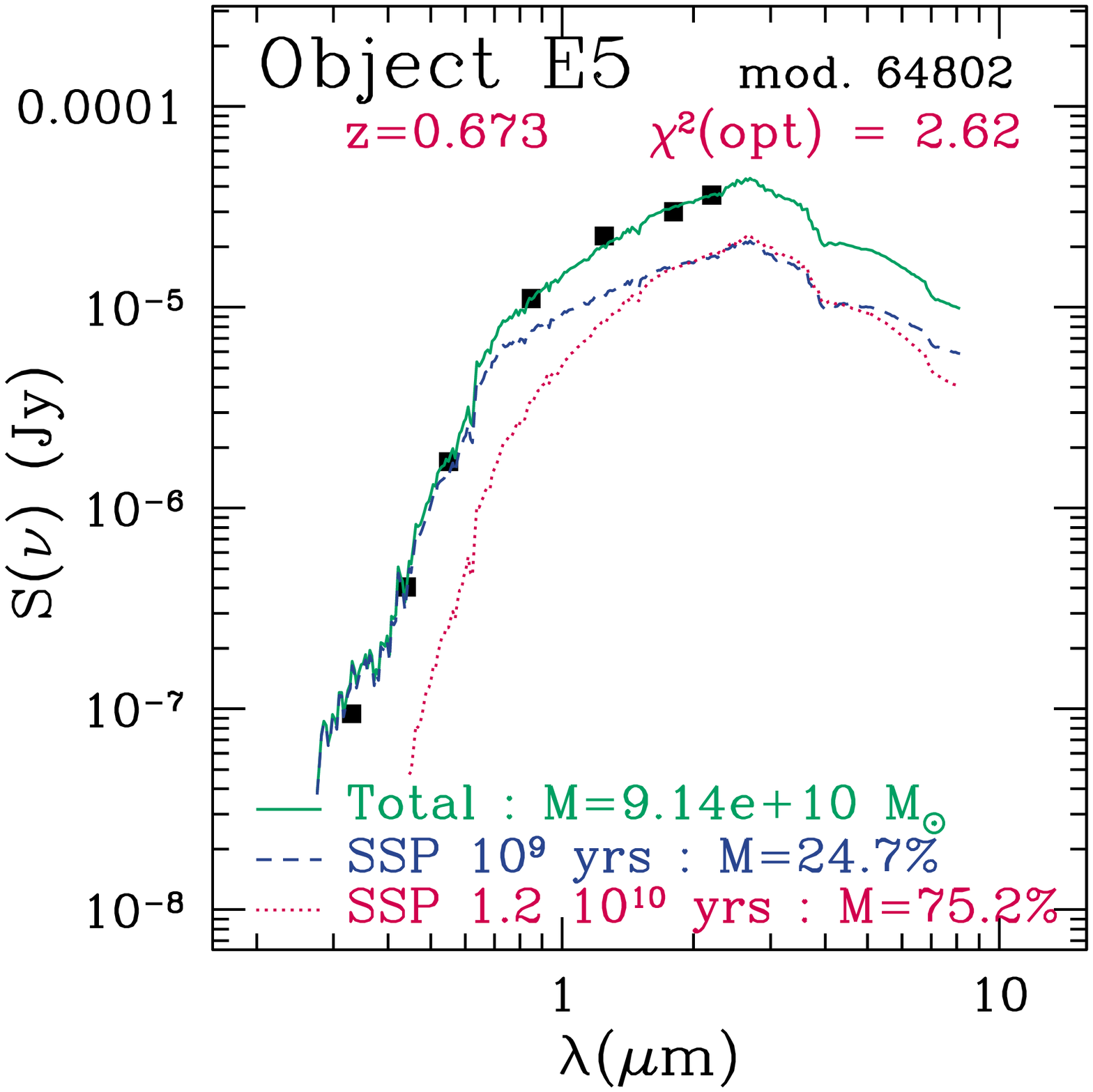}
\includegraphics[width=0.325\textwidth]{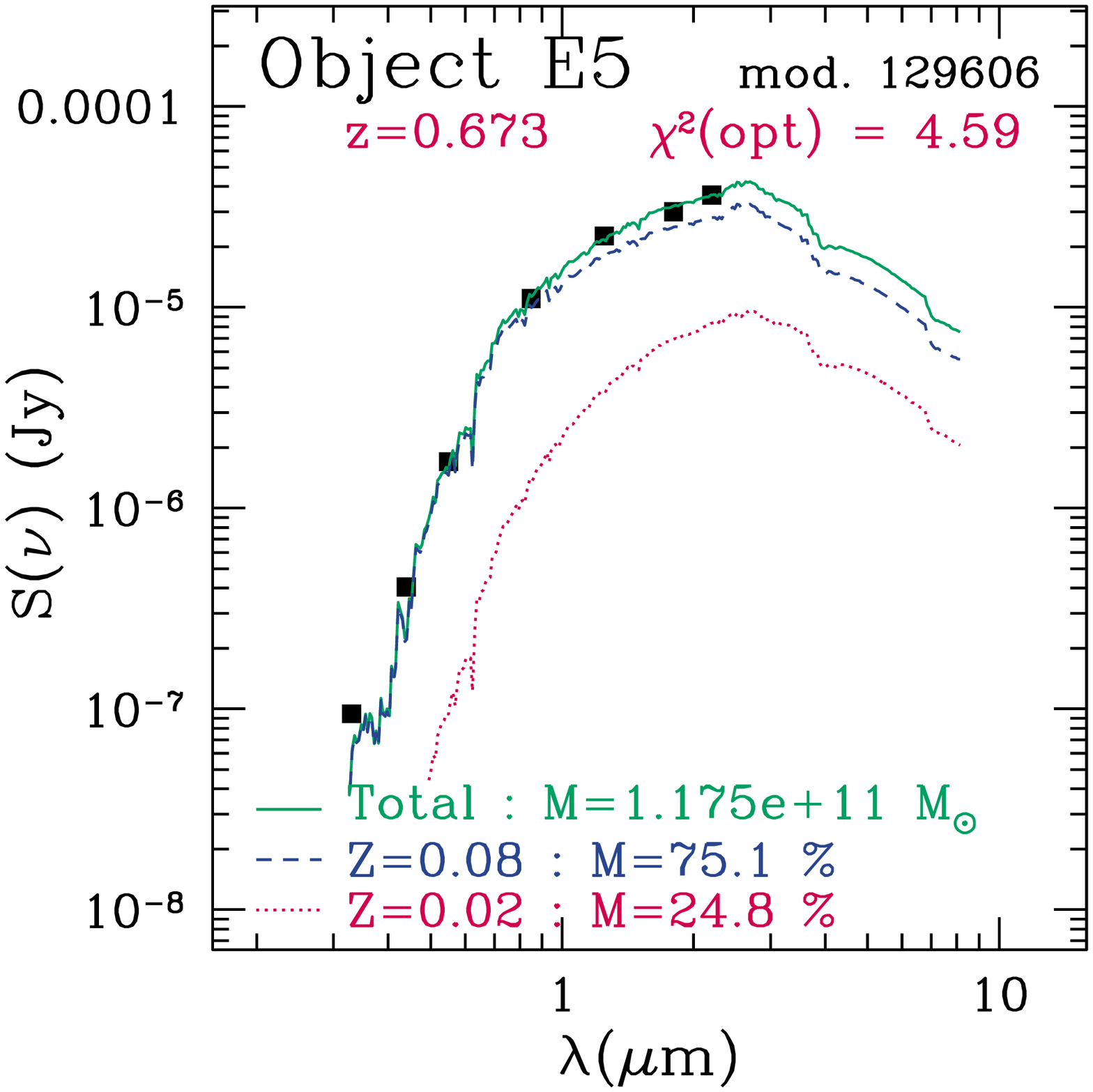}\\
\includegraphics[width=0.325\textwidth]{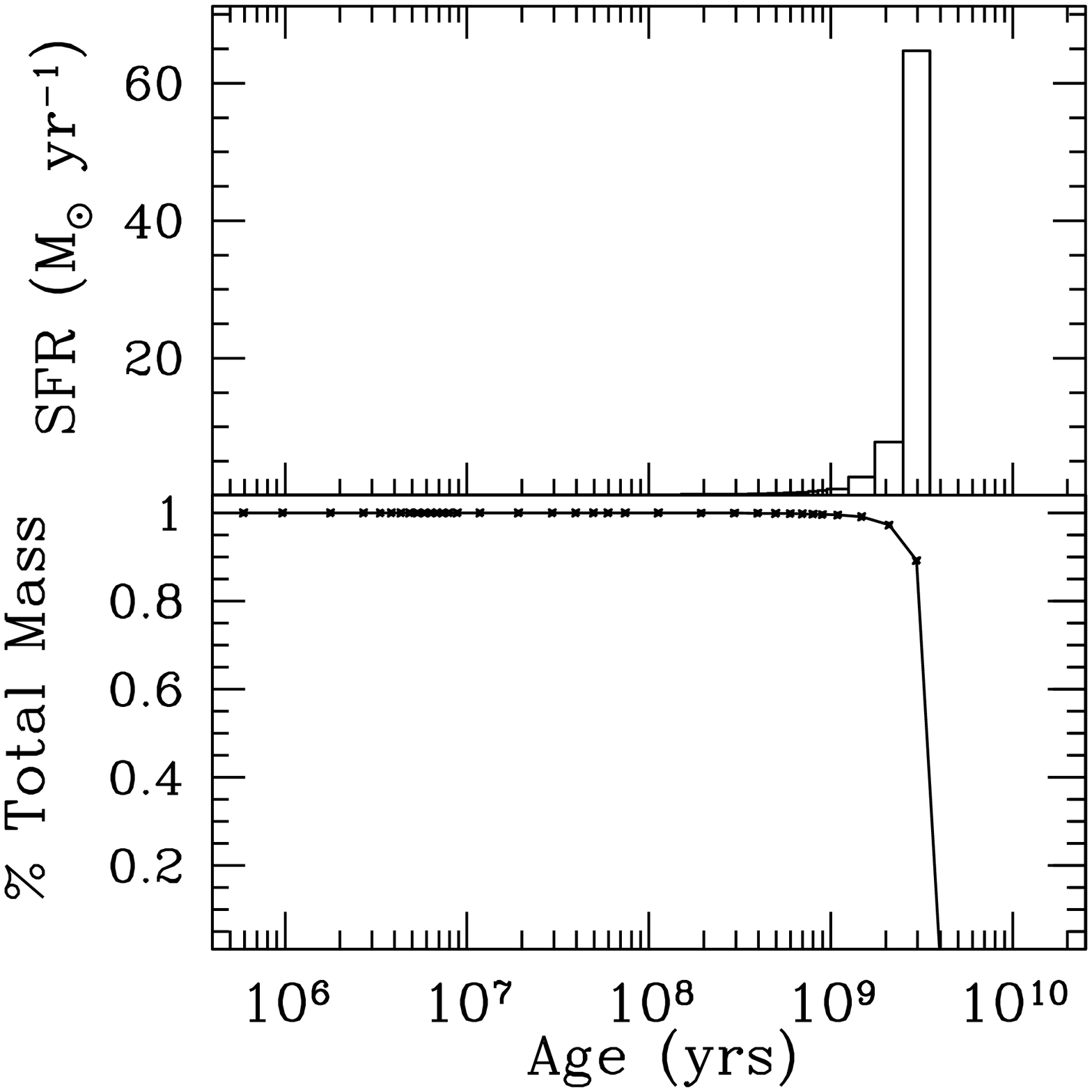}
\includegraphics[width=0.325\textwidth]{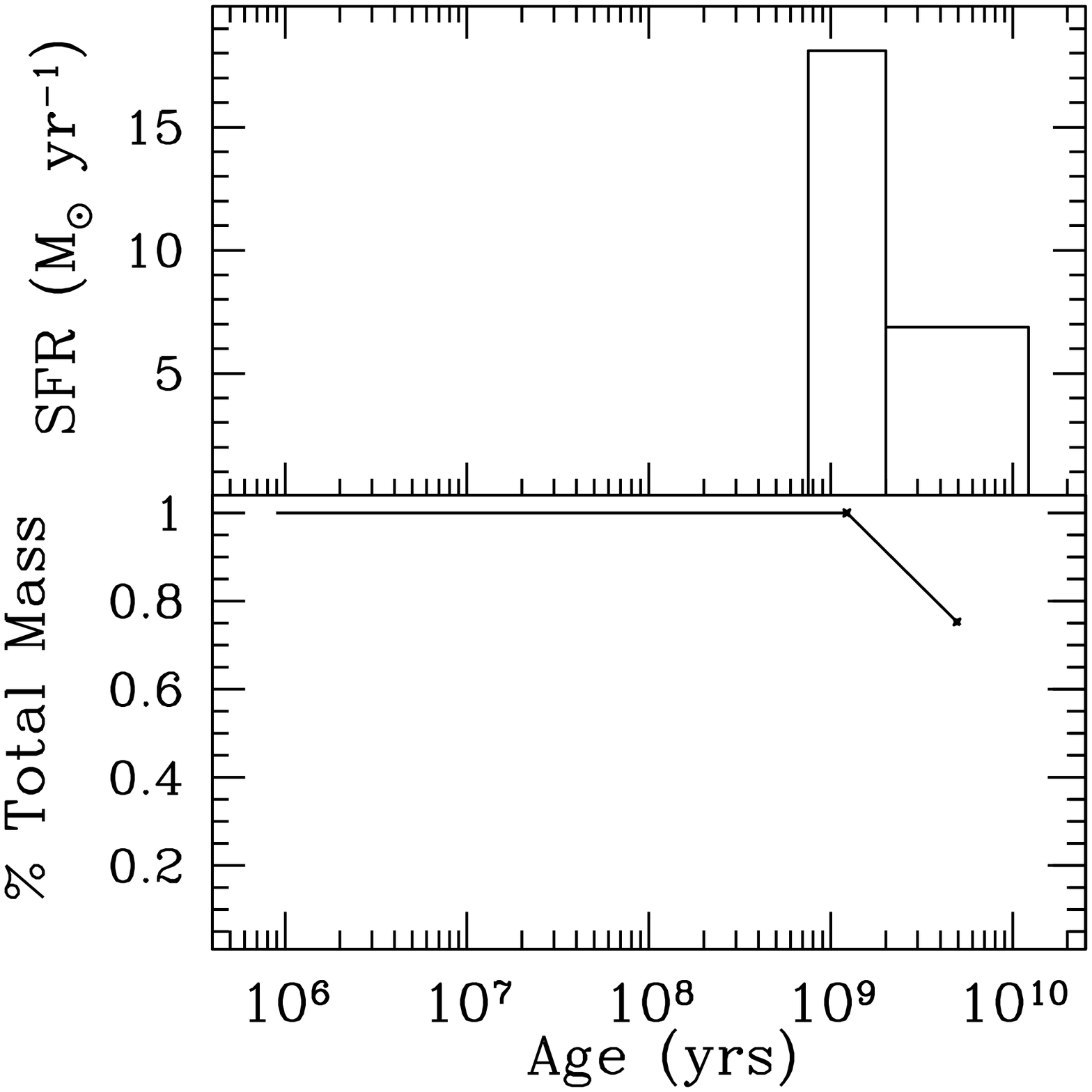}
\includegraphics[width=0.325\textwidth]{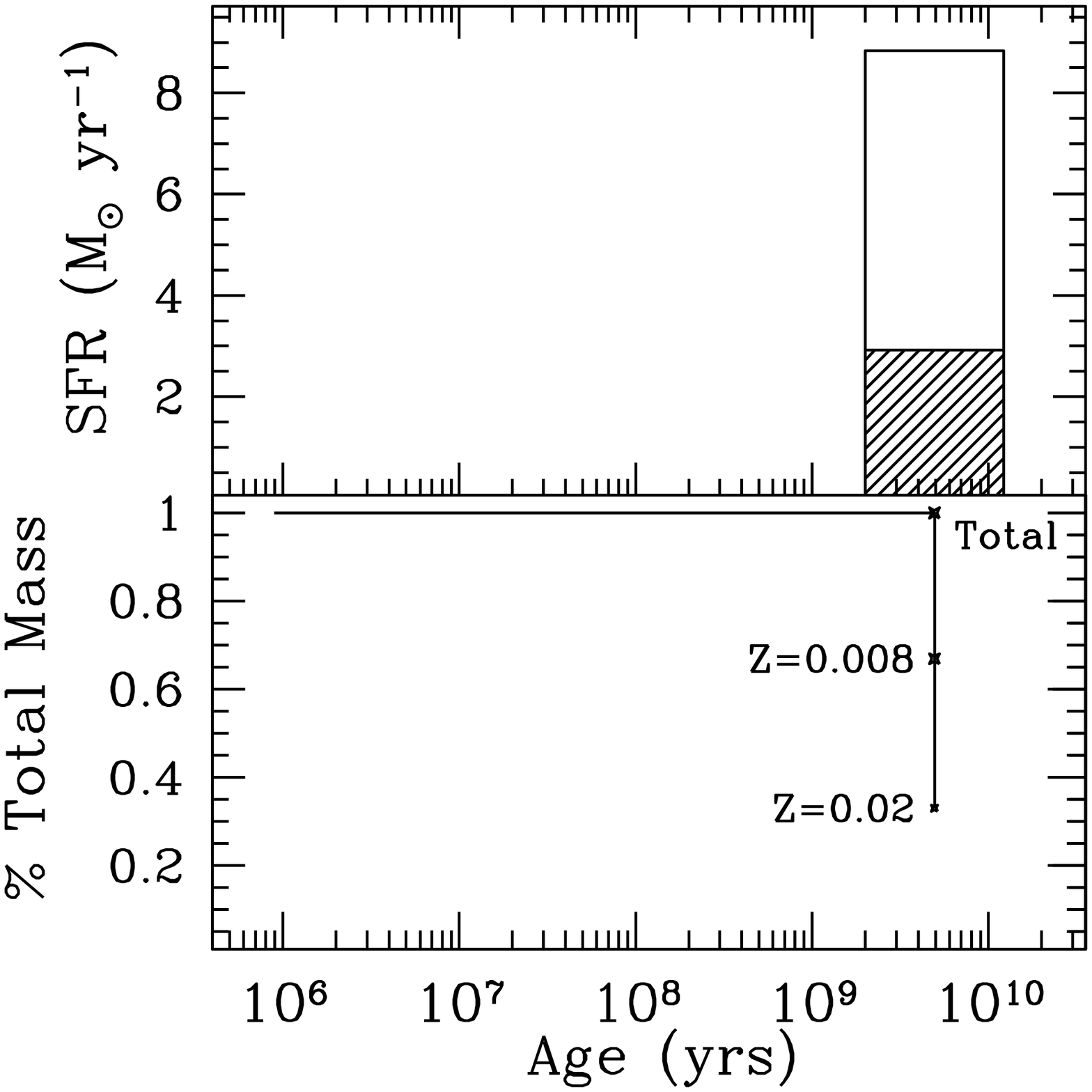}
\caption{Comparison of three different fits of the SED of source E5 in the
ellipticals HDFS sample by Rodighiero et al. (2001), obtained with the three
different adopted methods. {\em Left:} Analytic star formation history, given by a
Schmidt law; {\em center:} two single stellar populations (10$^9$ and 1.2 10$^{10}$
yrs old) with solar metallicity; {\em right:} two old populations (1.2 10$^{10}$
yrs) with different metallicities (Z=0.02, 0.008). {\bf TOP:} the best fit
models SEDs (solid line) have been split into two 
components: on the left into the contributions of SSPs younger than 10$^9$ yrs (dashed
line) and older (dotted line); at center into the contribution of the 10$^9$
yrs old SSP (dashed) and of the 1.2 10$^{10}$ yrs old one (dotted): on the right
into the contribution of the Z=0.008 SSP (dashed) and of the Z=0.02 one
(dotted).  {\bf BOTTOM:} star formation history and contribution of each population
to the total stellar mass assembled in the galaxy. In the bottom-right panel the
two old SSPs with different metallicities have been distinguished both in the
SFH plot (shaded histogram is the Z=0.02 SSP) and cumulative-mass diagram. 
The main contributions to the $\chi^2$ for the central and right solutions come
from the J and V data respectively. No extinction is plotted, since it was
assumed constant for all populations (see Table
\ref{tab:risultati_ellittiche}).}
\label{fig:e_5}
\end{figure*}

\begin{figure*}[!ht]
\centering
\includegraphics[width=0.97\textwidth]{dummy_f4.eps}
%\subfigure[E1: $z=0.512$, best fit $M=1.8\ 10^{10}\ M_\odot$.]{
%\includegraphics[width=0.245\textwidth]{berta_f4a.eps}
%\hspace{-0.3cm}
%\includegraphics[width=0.245\textwidth]{berta_f4b.eps}
%}
%\hspace{-0.4cm}
%\subfigure[E9: $z=0.515$, best fit $M=5.0\ 10^9\ M_\odot$.]{
%\includegraphics[width=0.245\textwidth]{berta_f4c.eps}
%\hspace{-0.3cm}
%\includegraphics[width=0.245\textwidth]{berta_f4d.eps}
%}\\
%\subfigure[E17: $z=1.25$, best fit $M=3.89\ 10^{11}\ M_\odot$.]{
%\includegraphics[width=0.245\textwidth]{berta_f4e.eps}
%\hspace{-0.3cm}
%\includegraphics[width=0.245\textwidth]{berta_f4f.eps}
%}
%\hspace{-0.4cm}
%\subfigure[E28: $z=0.564$, best fit $M=1.3\ 10^{10}\ M_\odot$.]{
%\includegraphics[width=0.245\textwidth]{berta_f4g.eps}
%\hspace{-0.3cm}
%\includegraphics[width=0.245\textwidth]{berta_f4h.eps}
%}
\caption{Best-fits spectral solutions for representative elliptical 
galaxies, based on the Schimdt star formation history model (see
Sect. \ref{par:ell_fit}). See also
Figs. \ref{fig:results_s40} and \ref{fig:e_5}.}
\label{fig:results_ell}
\end{figure*}

\subsection{Generalities on the spectro-photometric model}

To generate the synthetic spectra, we have developed a new automatic analysis
tool as an upgrade of the spectral synthesis code by  
Poggianti et al. (2001). We modified it on the computational side, in versatility 
and improved user-interface.

Each observed SED is modelled as a combination 
of a set of single stellar populations (SSP) of solar metallicity and different
age. Each SSP is meant to represent a formation episode of average constant  
star formation rate (SFR) over a suitable time period $\Delta$t.
Each single population is assumed to be extinguished by dust in a uniform 
screen according to the standard extinction law
of the diffuse medium in our Galaxy ($R_V=A_V/E(B-V)=3.1$, Cardelli et al. 1989;
see Poggianti et al. 2001 and Berta et al. 2003 for a discussion). 
The extinction value E(B--V) is allowed to vary from one stellar population to
another, and the total spectrum is built up by summing the extinguished spectral
energy distributions of all generations. 
However, in consideration of the fact that high extinction values are found only
for the young stellar populations embedded in thick molecular clouds and that
disc populations are on average affected by a moderate $A_V$ ($\le 1$, e.g. Kennicutt, 1992),
we have limited the E(B--V) values for the populations with ages $\geq 10^9\
yrs$ to be less than 0.30.  

All the SSP spectra have been computed with a Salpeter initial mass function (IMF)
between 0.15 and 120 $M_{\odot}$, adopting the Pickles (1998) spectral library extended 
with Kurucz (1993) atmosphere models.
The composite spectra, including both photospheric stellar and nebular (line and
continuum) emission,
have been obtained through the ionization code CLOUDY (Ferland, 1990).
This procedure provides a reliable description of single stellar generations up
to a wavelength of $\sim 5\ \mu$m (restframe). Beyond this wavelength, dust 
emission becomes typically no longer negligible.

The bulk of our analysis assumes SSPs of solar metallicity. However we can also
deal with synthetic spectra with different metallicities, i.e. one third and one
tenth solar, Z=0.008 and Z=0.002. As it will be subsequently discussed, the
estimate of the stellar mass content of a high-redshift galaxy barely depends on
the detailed assumption for metallicity.

The model SED is obtained by combining a number $N_{SSP}$ of different SSPs,
weighted by different values of SFR and different amounts of extinction: 
\begin{equation}
L^{unext}_{\lambda,i}=L^{SSP}_{\lambda,i}\cdot SFR_i\cdot \Delta t_i
\end{equation}
\begin{equation}
L^{mod}_\lambda=\sum_{i=1}^{N_{SSP}}\left( L^{unext}_{\lambda,i} \cdot 10^{-0.4
\frac{A_\lambda}{A_V}A_{V,i}} \right).
\end{equation}
The sum is intended over all the $N_{SSP}$ single stellar generations;
$L^{unext}_{\lambda,i}$ and $A_{V,i}$ are respectively the 
unextincted spectrum weighted by $SFR_i$ and the extinction value for the
$i$--th SSP, $L^{SSP}_{\lambda,i}$ is the unextinguished luminosity of the $i$--th
SSP normalized to 1 M$_\odot$, $\Delta t_i$ is the duration of the $i$--th SSP
and finally $\frac{A_\lambda}{A_V}$ is the standard extinction law ($R_V=3.1$).  

The result is corrected for cosmological dimming by applying 
the K--correction and compared to the
observed SED with:
\begin{equation}
S_\nu=\frac{L_\nu}{4\pi d_L^2}K(\nu,z)
\end{equation}
\begin{equation}
K(\nu,z)=(1+z)\frac{\int_{\nu_1}^{\nu_2}L[\nu(1+z)]T(\nu)d\nu}{\int_{\nu_1}^{\nu_2}L(\nu)T(\nu)d\nu}
\end{equation}
where $d_L$ is the luminosity distance at the given redshift (Carroll et al.,
1992) and $K(\nu,z)$ the K--correction factor; $T(\nu)$ is the filter
transmission function.

\subsection{Fitting the starburst galaxy SEDs}

The SEDs of the sources in our mid-IR sample have been modelled as a combination
of up to 10 SSPs, whose ages have been chosen by considering the evolutionary
timescales of the associated stars. We have then considered up to
four young solar-metallicity stellar generations (with ages of $10^6, 3\, 10^6, 8\, 10^6, 10^7$ yr), mostly
responsible for the ionizing photons producing the emission lines and for dust heating
and FIR emission; up to five intermediate--age populations
($5\, 10^7$, $10^8$, $3\, 10^8$, $5\, 10^8$, $10^9$ yr) with the
strongest Balmer absorption lines; finally for older generations of stars 
we have included a population modelled with a constant star formation rate (SFR)
between 2 and 12 Gyr before the observation. This latter choice is discussed
in section \ref{par:deg}, together with the effects of varying the SSP's metal content.

In the attempt of reducing the uncertainties in the mass estimate due to the contribution 
of young stars to the near-IR restframe SED, we have
considered the ISO 15 $\mu$m flux in addition to the UBVRIJHK photometric data.
The dust emission spectrum 
longwards of 5 $\mu$m (restframe) is modelled by assuming the observed M82 IR
SED and then comparing to the flux detected by ISO in the LW3 band (12 to 18
$\mu$m). 
The M82 IR template is adopted as a typical spectrum of luminous IR starbursts
(except some exceptional ultraluminous galaxies like Arp 220).
The predicted IR emission is calculated as the difference between the
unextinguished and the extinguished optical spectrum, assuming that all the flux
absorbed by dust is reprocessed and re--emitted longwards of 8 $\mu$m. The
M82 template is then rescaled in such a way that its bolometric emission between
8 and 1000 $\mu$m coincides with the dust reprocessed luminosity.
Finally, the properly scaled M82 SED 
is K-corrected and convolved with the LW3 filter reponse to compare with the
observed  LW3 flux.   

The M82 template was adopted to model the IR SED of the ISOCAM sources as a fairly conservative choice. 
Elbaz et al. (2002) have shown that tight nearly
linear correlations between the observed mid-IR emission and the bolometric FIR
luminosity of local IR galaxies hold over $\sim4$ orders of magnitude in
$L_{IR}$ (between 10$^9$ and 10$^{13}$ $L_\odot$), involving IRAS 12 $\mu$m and ISO LW2,
LW3 observed fluxes. 
M82, with $L_{IR}\simeq 3.3\ 10^{10} \ L_\odot$, is a prototypical galaxy,
representative of dusty starbursts and laying in the middle of the Elbaz et al. (2002) correlations.  

Objects not following these relations, e.g. the ULIRG
Arp 220, are characterized by a much more peaked FIR emission and dust
self-absorption.
Adopting such an extreme galaxy as IR template would produce higher bolometric
infrared luminosities  for a given 15 $\mu$m flux, and therefore require very
high --- often unphysical --- extinction values for those SSPs contributing to the FIR emission.
The estimate of stellar mass and its uncertainty (see Sect. \ref{par:deg}) is in
any case not significantly affected by this choice.

The exploration of the parameter space (see more details in Sect. 3.5 below) starts with a large number
(10) of stellar populations. 
However, after the minimization process and for each solution with acceptable
$\chi^2$, SSPs providing less than 1\% of the observed spectrum at each
wavelengths (after extinction has been applied) are identified and
excluded from further consideration.

The effective number of SSPs involved in the best-fit solution of each
examined object is reported in Table 
\ref{tab:results}. Tipically 3$-$4 stellar populations then contribute to the fit of
the observed SEDs, which in turn means a total of 6$-$8 parameters (SFR and
E[B--V] for each population).

The optimization process involves, at least at the beginning of the exploration,
a large number of populations/parameters to ensure that the 
most general description of galaxy spectra and the most conservative evaluation
is given of the uncertainties in the physical parameters.

\subsection{Fitting the elliptical SEDs}\label{par:ell_fit}

It is common wisdom that the properties of local ellipticals/S0s
may be reproduced by one suitable Simple Stellar Population with old age. 
However in some cases there are hints for the presence of {\em a)} a younger
population, representative of a partial renjuvenation episode occured in the
recent past; {\em b)} or a significant contribution from sub-solar
metallicity stars. Both effects result in a bluening of colors.

We have analysed the spheroidal galaxy population with various methods.
The first approach that we have considered consisted in adopting 
a simple analytical form of the star formation history (SFH) in E-type galaxies.
We have achieved this by combining 55 SSPs in an almost continuous age sequence
in which the stellar mass formed in each generation is obtained from:
\begin{equation}\label{eq:analy}   
SFR(t)=\left( \frac{t_F-t}{t_F}\right)^{n_1}\times \exp \left( -
\frac{t_F-t}{\tau_1 \cdot t_F} \right)
\end{equation}
where $t_F$ is the epoch of formation of the galaxy (i.e. when the first star
shone) and varies in the range $2-12$ Gyrs.
We assumed a constant small amount of extinction for all populations.
The relation \ref{eq:analy} describes the evolution of SFR in an open chemical evolution 
model obeying the Schmidt law. The power-law term represents the early growth of SFR.

The second method, meant to reproduce case {\em (a)}, consisted in the same free-form approach
already described for IR starbursts, but constrained in order to avoid high extinction
values and high FIR emission. The latter assumes that we have imaging data of enough
quality to ensure that the object under scrutiny is a {\sl bona-fide} spheroidal
galaxy not affected by extinction features as revealed by dust lanes, absorption
bands, etc.       For the typical E/S0s in our sample, best-fits are obtained 
with the combination of just two SSPs, one of $1.2\ 10^{10}$ yrs and the other 
of $10^9$ yrs, plus a younger SSP needed for the blue ellipticals.

Particularly to fit the case {\em (b)} sub-population, we have considered a third choice: 
a combination of two old single stellar populations, both of them 1.2 10$^{10}$ yrs old,
having different values of metallicity (solar Z=0.02 and sub-solar Z=0.008).

As a final, check we have also considered the most general case of a completely free-form
approach, as that mentioned above to fit the dusty star-forming class, but with 
an upper limit imposed on the mid-IR flux (as might be provided by
forthcoming SIRTF multiband surveys) to constrain to amount of dust re-radiation
and optical extinction.

\subsection{Fitting the Ly-break galaxy SEDs}

\begin{figure*}[!ht]
\centering
\includegraphics[width=0.97\textwidth]{dummy_f5.eps}
\caption{SEDs fits of some Lyman-break galaxies. {\em Left}: best fit solutions
for sources Ly2, Ly3 and Ly5, as obtained involving only young SSPs. {\em
Right}: consequences of introducing the $\ge10^9$ yrs old populations on the fit of
source Ly1. The relative contribution of old stars to the total assembly of
stellar mass increases from top to bottom: the fit accuracy does not change, but
in the restframe $1-2 \ \mu$m domain the model becomes significatively steeper 
(see also right panel of figure \ref{fig:s40_sirtf} and text for details).} 
\label{fig:ris_ly} 
\end{figure*}

The spectral energy distributions of the Ly-break galaxies have been
fitted assuming a discrete number of stellar populations as for the IR
starbursts, but since at redshift $z\sim 2.5$ the age of the universe for the
assumed cosmology is roughly 2.5 Gyrs, only SSPs up to 1 or 2 10$^9$ years 
--- depending of $z$ --- have been considered. 

As for the starburst case, only stellar populations contributing to the visible
and IR light by more than 1\% were involved in the fit. The typical number of
SSPs used is 2$-$3 for Lyman-break galaxies (see Table \ref{tab:results_ly}).

\subsection{Parameter space exploration with the Adaptive Simulated Annealing (ASA)}
\label{par:asa}

For each galaxy
the code randomly generates a wide set of models and compares them to the
observed fluxes, by minimizing the {\em Merit Function}, which tests
the accuracy of the given fit:
\begin{equation}
\chi^2=\sum^N_{i=1}\left( \frac{S^{obs}_i-S^{mod}_i}{\sigma^{obs}_i} \right)^2,
\end{equation}
where $S^{obs}_i$ and $S^{mod}_i$ are the observed and synthetic fluxes in the
$i$--th photometric band, $\sigma^{obs}_i$ is the uncertainty on the observed
flux and $N$ the number of constraints.  

In the most general case there will not exist a unique minimum of the Merit
Function, but rather a large number of relative minima.   
The exploration of the $N$--dimensional parameter space has been performed
adopting the {\em Adaptive Simulated Annealing} algorithm (ASA, Ingber 1989,
2001), which is very well suited to search for a solution as close as possible to the
absolute minimum, avoiding local minima.   

The heart of the {\em Simulated Annealing} method (Ingber et al., 1989, 2001)
is the thermodynamic analogy with cooling metals:
at high temperatures each molecule moves with respect to the others, but
when $T$ slowly decreases its mobility is gradually lost and matter reaches
a pure crystalline configuration. 
If the cooling process is slow enough, metals can spontaneously reach this {\it
minimum energy} state. 

The function to be optimized can be considered similar to the total energy of the 
system. The Boltzmann probability distribution
%$$
\begin{equation}
P_E \sim e^{-\frac{E}{KT}}
\end{equation}
%$$
shows that even at thermal equilibrium -- with temperature $T$ --
a non-zero probability to find the system in an energy state higher
than $KT$ does exist. 
Consequently it is possible for the system
to move from the local energetic minimum to an excited state and subsequently
evolve to an energy smaller than at start. 

Following this analogy, the optimization process starts from a given state in
the $N$-dimentional parameter space and then evolves from $E_1$ to $E_2$;
the probability of this transition is $P_{12} = e^{-\frac{E_2-E_1}{KT}}$.
If $E_1>E_2$, then $P>1$: the system always moves towards lower energy states,
even if a non-zero probability to increase its energy does exist.\\
On the other hand, 
small values of the temperature and large energetic increments imply low
probabilities for the system to evolve uphill.  

In order to converge, at each $i$-th iteration, the temperature of the
system is {\em adapted} according to the relation
$T(i+1) = R_T \cdot T(i)$.
The adopted value for $R_T$ is 0.85, as suggested by Corana et al. (1987). 
As $T$ declines, moves to excited energy states are suppressed and the algorithm
focuses on the most promising area for optimization.
On the other hand, when a new deep minimum is found, these ranges
are increased, in order to avoid the system to lag on it. 

Minimization ends when the $\chi^2$ value decreases by less than the {\em
Tolerance} set by the user. 
Further informations on the ASA algorithm can be found at the URL {\tt
http://www.ingber.com/}.

\section{Results}\label{par:results}

For each of the sample sources, our simulations generate a very large
number of models, each one based on a different set of physical parameters.
For a typical number of computed solutions ($\sim 100,000$), the $\chi^2$ exploration
takes between $\sim$20 and 45 minutes on a 1.7GHz CPU PC -- depending on the
total number of SSPs, the adopted SFH and convergence criteria. Acceptable
solutions (with reduced $\chi_0^2\sim 1$) are found for most of the sample
galaxies. 
In the case of unacceptable fits, it may be that the uncertainties in the
measured fluxes were underestimated, or that the distribution of the errors on
parameters is not gaussian (Papovich et al., 2001). Finally, the SSP spectra may
have intrinsic modellistic uncertainties.

\subsection{Mid-IR selected starburst galaxies}

Figure \ref{fig:results_s40} illustrates the comparison of the
synthetic SEDs to the observed datapoints and plots of the main physical
parameters, for a representative galaxy.  The left panel in the figure
shows the best-fits to the SEDs. The solid line is the sum of contributions from
all SSPs; the dashed line is the sum of all the young SSPs (age $< 10^9$ yrs),
while the dotted line represents the contribution of older populations (age $\ge
10^9$ yrs); the dashed thick line longwards of 5 $\mu$m is the redshifted M82
template; the square black dots are the observed photometric datapoints, the red
triangle at $\lambda\sim 15 \ \mu$m is the fit to the ISO LW3 datapoint. 
The identification number of the model (for the given source) and the $\chi^2$
value based on the optical/NIR fit only (LW3 datapoint excluded) are also reported
in the plots. 

The right panels of Fig. \ref{fig:results_s40} show --- 
for the various SSPs making up the best-fit model, whose 
age intervals are indicated in the x-axis --- trends
of the SFR (top histograms, y--scale on the left end), E(B--V) (top panel, filled squares, 
y--scale on the right end),
the ratio of the FIR flux contributed by the SSP to the total flux in
the V band ($F_{FIR,SSP}/F_{V,TOT}$, middle panel), and the cumulative distribution 
of the stellar mass as a function of SSP's age (bottom panel). 
The graph in the middle panel, in particular, details the contribution of each
SSPs to the FIR bolometric emission of the galaxy, taking into account the effect
of dust extinction in the optical light. The bottom plot shows that, typically,
the youngest stars contribute to the bulk of the FIR bolometric emission of
these sources, but provide almost no contribution to the optically luminous mass
in the galaxy.  
Results for additional dusty starbursts in Hubble Deep Field South
may be found in Franceschini et al. (2003).

\subsection{Degeneracies in the mass estimates for IR-selected dusty starbursts}\label{par:deg}

The left panel of Figure \ref{fig:contours_s40} shows plots
of $\chi^2$ contours as a function of stellar mass $M$ and FIR luminosity
$L_{FIR}$, based on fits to the optical/NIR data only (UBVRIJHK photometry). 
The $\chi^2$ contours correspond to 1, 2, 3$\sigma$ uncertainties.
Note that even the darker 1$\sigma$
contours are extended, showing substantial degeneracies in the solutions. 

The horizontal lines in Fig. \ref{fig:contours_s40} (left panel)
mark the measured FIR bolometric (8$-$1000 $\mu$m)
luminosity and its uncertainty based on the fits to the observed LW3 flux.
In only a fraction of the high-z galaxies the FIR constraint is effective in 
reducing the uncertainties in the stellar mass: this happens in particular
for the least active galaxies (those with the lowest mid-IR excess and SFR compared 
with the optical SED). Source S27 ($z=0.58$) is one such case: imposing the LW3
flux constraint, the stellar mass range is reduced by more than a factor of two. 

On the other hand S40, an ultraluminous galaxy at z=1.27, is a typical case
for which the FIR information is barely useful to constrain the mass.
This fact is due to the presence of highly obscured young stellar populations in
the most active starbursts, which give a significant contribution to the 
NIR and FIR luminosity, but are not visible in the optical/NIR, because of their 
low M/L ratio, and their contribution to the stellar mass budget is hard to constrain. 
For such complex situations, further mid-IR data from SIRTF will be critical,
as discussed in Sect. \ref{SIRTF} below.
 
As mentioned in Sect. 3.2, we have reproduced the contribution by old stellar
populations with spectra obtained assuming constant SF between
2 and 12 Gyrs, since the single SSP spectra in this age range are all very similar
and their M/L ratio varies smoothly. This allows us to reduce the number of free 
parameters, when compared to a more detailed SFH.
To test this procedure, we have ran different simulations on several sources, detailing the SFH between 2
and 12 Gyrs into 5 different SSPs (2, 4, 7, 10 and 12 Gyrs old). Other
simulations obtained by simply substituting the long-SF model with single 2 and 12
Gyrs SSPs have been run as well, in order to test the more extreme cases of
masses assembled with the lowest and the highest M/L values in the 2-12 Gyrs range. 
The most massive and the lightest models obtained in this way may differ by a factor of
$\sim 2$ in mass. However, this analysis showed that the uncertainty $\Delta$M in stellar mass -- due to different adopted models of the old population -- is in fact completely contained within those of 
the global model fit. Indeed the $\Delta$M due to the presence of young stellar
populations dominates the global mass uncertainty.

Although we have based our analysis on solar-metallicity SSPs, we have performed a systematic analysis of how variations in metallicity
may affect our results, taking into account that e.g. changing 
it from solar to one third solar would be equivalent to change the SSP age from
10$^9$ to $12\times 10^{9}$ yrs (the age-metallicity degeneracy). In
principle one would expect a corresponding significant difference in the M/L ratio, hence on the estimated stellar mass.
In fact our analysis has shown that a kind of compensations holds, such that
changing the stellar metallicity requires a substantial modification of the SF history to keep an acceptable spectral fit. The result is that the mass estimate does not significantly depend on the assumed metal content, and the same applies for the mass uncertainty range.
For star-forming galaxies the overwhelming uncertainty factor is instead the incidence of the young stellar population. 

\begin{figure*}[!ht]
\centering
\rotatebox{-90}{
\includegraphics[height=0.9\textwidth,width=0.4\textheight]{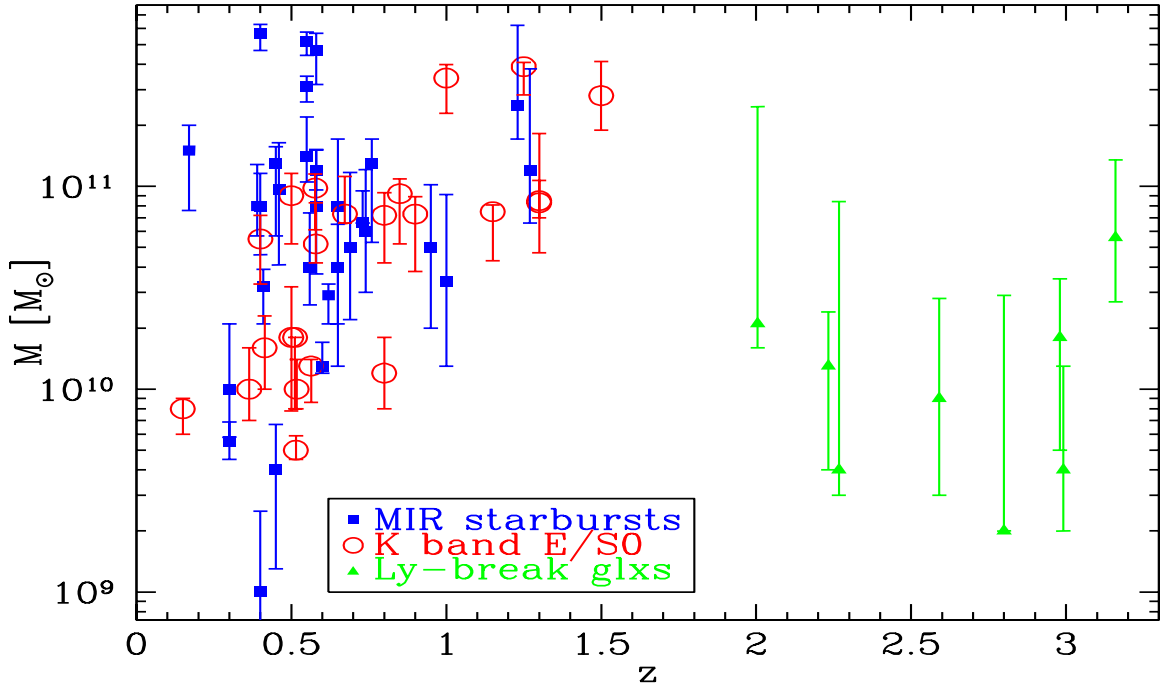}}
\caption{Comparison of stellar masses between mid-IR selected starburst (filled
squares), K-band selected E/S0 (open circles) and Ly-break galaxies
(triangles), as a function of redshift. With the exception of very massive MIR
sources belonging to the HDFS well-known group at $z\simeq$0.6, the stellar
masses we find for starbursts and E/S0 are comparable; the small sample of
Ly-break galaxies seems to consist of less massive objects than those observed
at lower redshift.} 
\label{fig:z_mass} 
\end{figure*}

The stellar mass estimates are reported in Table \ref{tab:results}. 
We plot in the right panel of Fig. \ref{fig:contours_s40}  the
$\chi^2$ values as a function of mass for acceptable models,
each dot representing a different solution. 
Columns 8 and 9 of Table \ref{tab:results} report the estimated uncertainties $\Delta$M at 1 and 2$\sigma$ levels. 
The typical uncertainty in the mass values for these IR starbursts 
is a factor of 2$-$3, but may be as high as 5 in a few cases. 
Figure \ref{fig:z_mass} shows the estimated masses against redshift.

\subsection{Elliptical galaxies}

Franceschini et al. (1998) and Rodighiero et al. (2001) found that populations of
morphologically-selected spheroidal galaxies in the HDFs are characterized by a
wide variety of integrated colors, including typical ellipticals with very red
SEDs (B--I$>$2.5 for objects at $z\leq 1$ and V--J$>$2.5 for $z\geq 1$),
but also many bluer objects.
The b/r flag in the third column of Table \ref{tab:risultati_ellittiche}
identifies blue and red objects in our sample.

As previously mentioned, the SEDs of elliptical and S0 galaxies have been fit
alternatively by assuming an analytic star formation history, or by means of
the free-form approach, or the sub-solar metallicity model.
The SEDs of the red subsample are equally well-fitted by all three, 
and the corresponding estimated values of the stellar masses agree very well 
within the relatively small uncertainties. For these sources, the mass range 
reported in Table \ref{tab:risultati_ellittiche} includes the results of the 
three fits, while the best fit value refers to the analytic SFH tecnique.

Figure \ref{fig:e_5} shows best fits for the three methods for source E5: all
three yield equally good fits, although the analytic SFH gives the
lowest $\chi^2$; the lower panels of Fig. \ref{fig:e_5} report the star formation
histories and the cumulative mass distribution in the three cases.
Table \ref{tab:risultati_ellittiche} summarizes our results; the estimated
stellar masses are in fair agreement with those by Rodighiero et al. (2001).

For the blue E/S0 galaxy population, the simple combination of 2 old populations 
cannot reproduce the observed U, B (and sometimes V) observed fluxes, while good 
solutions can be found by assuming some ongoing star formation (either within the 
analytic model, see Fig. \ref{fig:results_ell},
or by adding a third young SSP in the discrete SFH).

Finally we have investigated the implications of analysing the SEDs for this
class of galaxies with the most general free-form spectro-photometric tool based 
on the multi-SSP approach and including stronger extinction effects. The only 
assumption in this experiment is that we have an upper limit to the mid- and
far-IR flux to constrain possible dust re-radiation
(we put a limit of 70 $\mu$Jy at 24 $\mu$m, a value easily achievable with 
SIRTF observations). For red ellipticals this increased freedom in the spectral
analysis has no effect in modifying the stellar mass estimate, while for
blue ellipticals the uncertainty range increases by typically 20\%, with
respect to the case when we can exclude {\sl a-priory} important extinction
e.g. based on imaging data.

\subsection{Comparison of stellar masses for moderate-z starbursts and E/S0
and high-z Ly-break galaxies}

Figure \ref{fig:ris_ly} reports the best fit solutions for four of the Lyman-break
sample galaxies based on models including only populations younger than 10$^9$ yrs. 
 The right-hand panels, top to bottom, of the figure 
illustrate the consequence of introducing an increasing fraction of 1-2 Gyr-old
SSP in the spectrum of source Ly1. The overall fit to the observed optical-NIR
data and $\chi^2$ values do not change, 
but the synthetic SED longwards of the K-band is significantly modified. 
As a consequence, without constraints on the rest-frame near-IR spectrum where
the bulk of the emission by low-mass stars is expected, the uncertainties in
the photometric mass estimates are very large, typically 1 order of magnitude 
(see numerical details in Table \ref{tab:results_ly}). 

Figure \ref{fig:z_mass} plots the estimated baryonic masses for our three
galaxy samples as a function of redshift.
On one side, this provides evidence that the IR-selected starbursts and
the K-band selected ellipticals and S0 galaxies seem to be comparably massive
systems. 
Note however that the small sampled volume may imply important cosmic variance effects (e.g. some structures are evident in the HDF South samples around z=0.5-0.6, including galaxies of huge stellar masses). 

Fig. \ref{fig:z_mass} shows that, in spite of the large mass uncertainties, our small illustrative sample of Lyman-break galaxies seem to
correspond to a galaxy population significantly less massive ($M<$ a few
$10^{10}M_\odot$) then those observed at lower redshifts (for which typically
$M>$ several $10^{10}M_\odot$, except for z$<$0.7 where low-mass galaxies
become detectable) over comparable sky areas.  
Similar results were found by Papovich et al. (2001) on the sample of HDFN faint
Lyman-break galaxies, with a median stellar mass of $M\sim6\, 10^9$ M$_\odot$ and by Shapley et
al. (2001) on brighter Ly-break galaxies (with a slightly higher median mass).
Although not conclusive, due to the lack of statistics and completeness, these results 
indicate substantial stellar build-up to happen at $z\sim 1$ to 2 in the field population.
This seems consistent with the steady increase of galaxy's M/L from z=3 to z=0 and the comoving stellar
mass density increasing by a factor of $\sim 10$ between z=2 and z=1 claimed by Dickinson et al. (2003).

\section{Perspectives for the SIRTF mission}\label{SIRTF}

\begin{figure*}[!ht]
\centering
\includegraphics[width=0.46\textwidth]{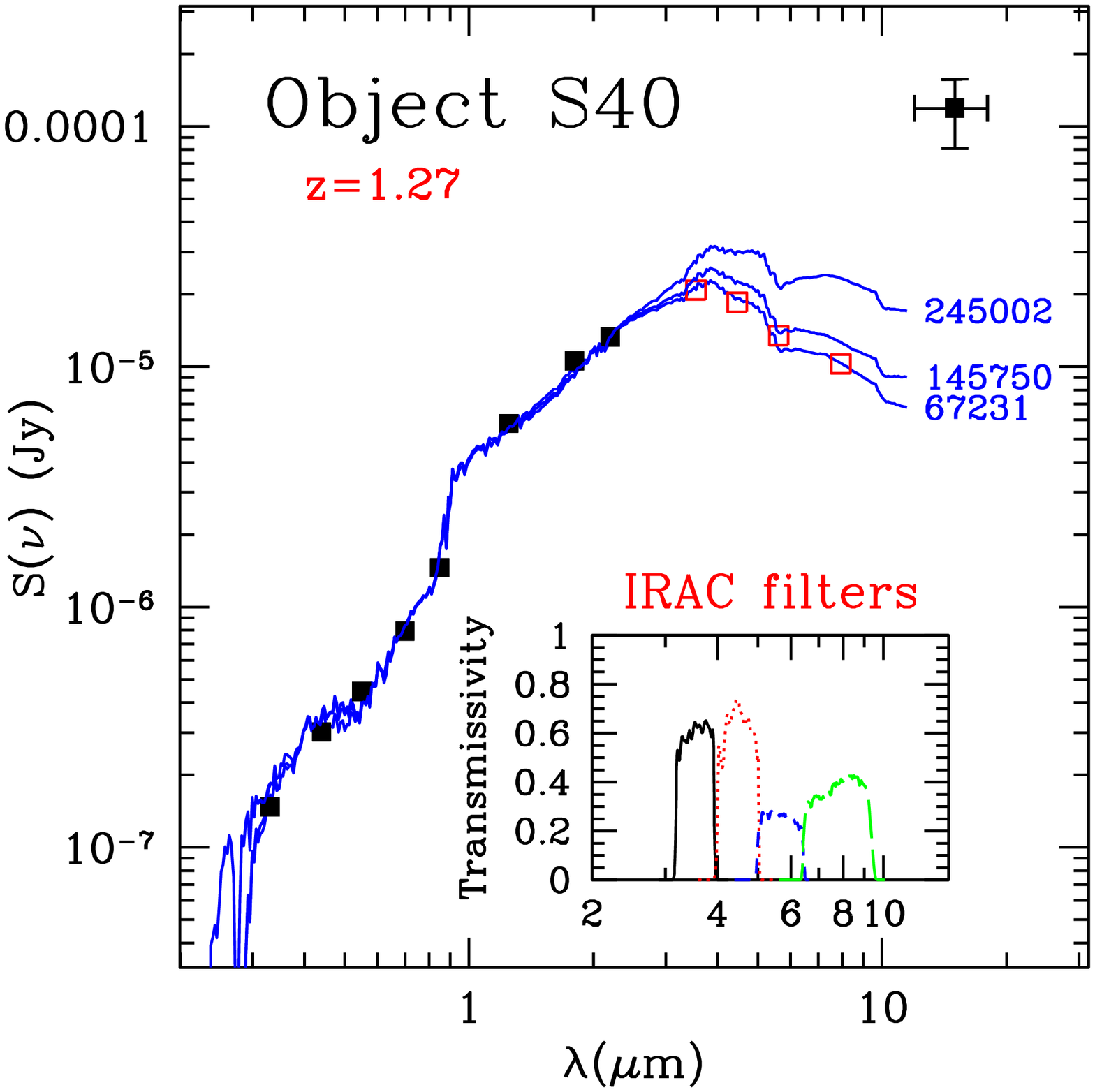}
\includegraphics[width=0.46\textwidth]{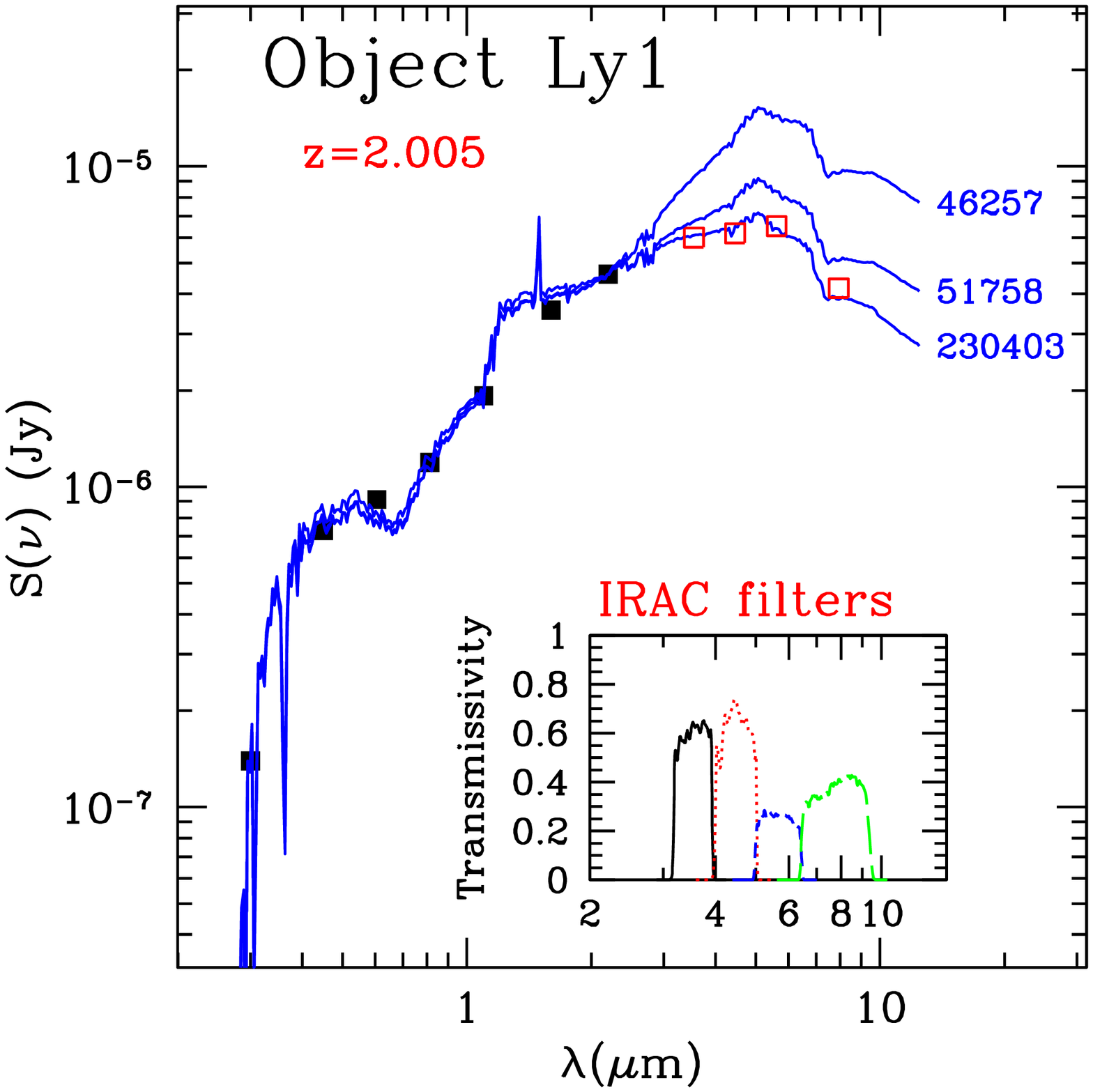}
\caption{Simulations of SIRTF/IRAC observations to constrain the estimate of stellar
masses in starbursts and Lyman-break galaxies. {\em Left panel}: three different
solutions for ISO/LW3 source S40: all three fits reproduce comparatively
well both the optical data and the LW3 observed flux (even if not displayed for
plotting reasons). SIRTF synthetic data (open squares) have been obtained
simulating an IRAC observation of the best fit model. {\em Right panel}: the
three different fits to optical-NIR data of source Ly1 already discussed in
Fig. \ref{fig:ris_ly}. IRAC synthetic fluxes have been computed for the best
fit solution, which is the youngest and least massive among the three considered.
In both panels the three models differ in their star formation history: the
reddest are dominated by old populations more than the others. Small boxes
show the transmissivity curves of the four IRAC filters; the numbers of the
models refer to those appearing in Figs. \ref{fig:results_s40} and
\ref{fig:ris_ly}.} 
\label{fig:s40_sirtf}
\end{figure*}

The newly operative NASA's Space InfraRed Telescope Facility (SIRTF) includes
 a sensitive camera (IRAC) to observe high-z galaxies in complementary
long-wavelength channels between 3 and 10 $\mu$m, hence providing us with a unique
opportunity for the analysis of stellar masses at intermediate and high
(0.5$-$3) redshifts, by sampling the critical restframe near-IR domain. 

We have simulated SIRTF/IRAC observations based on the best fit models of some
sources in the selected starburst and Lyman-break galaxy samples, to attempt
to quantify how such observations will improve the mass estimates.

\subsection{Constraining the mass of active starbursts at $z=0.5-1.5$}

In the case of the IR-selected starbursts, we have found
uncertainties in the stellar mass estimate based on the optical/NIR+LW3 SED
of typically factors of 3--4. 
SIRTF/IRAC, observing at 3.6, 4.5, 5.8 and 8.0 $\mu$m, will significantly improve
upon this.
Figure \ref{fig:s40_sirtf} illustrates this with an example. The open squares are the expected fluxes for galaxy S40 in the SIRTF/IRAC bands,
assuming our best-fit spectral solution. Some other solutions,
characterized by different SFHs, are also plotted. 
The case of S40 is representative of the high redshift population, 
while for intermediate $z$ objects (e.g. S53 or S55, $z\sim0.5-0.7$) the
improvement may be less significant. 

We have simulated a SIRTF/IRAC observation of twelve randomly-selected 
sources in the LW3 dusty starburst sample. We have
convolved the best-fit optical/NIR SED with the 4 IRAC band
transmission curves. 
Then we have run our code of spectral synthesis on the 12 objects, assuming
IRAC simulated fluxes with a 10\% photometric errors.
As a result, the estimated stellar--mass range is expected to reduce 
to a typical factor of
$\sim 2$, as reported in the last column of Table \ref{tab:results}.

\subsection{Detecting NIR restframe emission of $z=2-3$ Lyman-break galaxies}

Lying at redshifts $z>2$, the Lyman-break galaxies in our sample will
benefit by IRAC observations much more than intermediate and low-z galaxies.
In this redshift range, IRAC $3-8\ \mu$m data will sample the restframe near-IR
emission of galaxies, hence observing the light emitted by low-mass stars
dominating the galaxy stellar content.  

We simulated IRAC observations based on our best fit
solutions. Then we have applied our spectral synthesis code using 11 datapoints, 
namely the already available UBVIJHK 
photometric data and the synthetic IRAC 3.6, 4.5, 5.8 and 8.0 $\mu$m fluxes. 
Figure \ref{fig:s40_sirtf} (right-hand panel) shows how IRAC
observations will constrain the three solutions shown in the right panel of figure
\ref{fig:ris_ly}.

The results of these simulations are summarized in the last column of Table
\ref{tab:results_ly}: the mass uncertainty very significantly reduces to a
factor of $\sim 2-3$.

\section{Summary}

We have presented a new spectrophotometric synthesis code for the integrated 
light of distant and high-redshift galaxies. 
This tool is aimed, in perspective, at investigating various fundamental 
physical parameters, like the age of the dominant stellar populations,
extinction, SF history, ongoing star-formation rate, etc. We have concentrated in this paper to study in particular 
the ability of the code to constrain the stellar mass from SEDs fitting.
The tool has been tested on a sample of luminous infrared starbursts detected by ISO
in the Hubble Deep Field South at 15 $\mu$m (Franceschini et al., 2003), on a
HDFS K-band selected sample of ellipticals and S0s (Rodighiero et al. 2001),
and on a set of the Lyman-break galaxies in the HDF South from Papovich et al. (2001).

We have fitted the optical-NIR SEDs (from U to K bands) of the mid-IR starbursts 
by assuming a discrete star formation history between 1.2\,10$^{10}$ years ago
and today, by means of the combination of up to 10 single stellar populations.
Each SSP was weighted by different SFR's and absorbed by different amounts of
dust. For a best exploitation of the available constraints we also included in
the analysis the 15-$\mu$m ISO flux, to assess the contribution of 
extinguished young stars to the emitted spectrum. 
Based on these data, our estimated mass uncertainties range from a factor $\sim$2 
to occasionally a factor of 5. In such dust-obscured starburst galaxies, 
substantial uncertainty in the mass estimate comes from the still possible existence
of strongly extinguished young stars with low M/L ratios, contributing to the
far-IR flux but undetectable in the rest-frame optical.

We have also analysed a sample of morphologically-selected E/S0 galaxies,
at z$<$1.5.  Their SEDs have been reproduced
as the combination of stellar populations with solar metallicities, or by 
combining SSPs of different metallicities (Z=0.02 and Z=0.008). 
Additional solutions have been sought with a parameterized continuous sequence 
of stellar ages. 
The SEDs of spheroidal galaxies with red colors are well fitted by combining 
intermediate and old populations only, while for ellipticals with relatively 
blue colors (the majority in our complete sample) acceptable solutions are 
found only by assuming some some recent or even ongoing star formation.
The three different methods provide results well consistent with each other.

For Lyman-break galaxies the available rest-frame optical-UV data are much less
constraining.
The consequences of assuming relatively old ($\sim 10^9$ yrs) stellar
populations in these systems have been discussed and it has been 
found that the lack of long-wavelength data imply uncertainties in the mass
estimate of up to a factor of $\simeq 10$. 
In spite of these large uncertainties, there are indications that 
typical Lyman-break galaxies correspond to a galaxy population significantly
less massive then those observed at lower redshifts, possibly indicating
substantial stellar build-up to ocurring at $z\sim 1$ to 2 in the field galaxy
population. 

The situation is expected to significantly improve, particularly for higher-z
galaxies, with the forthcoming SIRTF/IRAC observations. Our simulations show
that such an extension of the wavelengths dynamic range will reduce the mass
uncertainties to factors of 2$-$3 for various classes of galaxies up to
z$\sim$3. This promises to be good enough for statistically reliable
determinations of the galaxy evolutionary mass functions.

\begin{acknowledgements}
We wish to thank M. Rowan-Robinson for useful and interesting discussions on
SEDs fitting and degeneracies. SB work was supported by ASI research grant
no. I/R/062/02. We are grateful to the anonymous referee for his very useful 
comments.
\end{acknowledgements}

% tabelle con masse e ranges
\begin{table*}[!ht]
\centering
\begin{tabular}{l c c c c c c c c}
\hline
\hline
Obj & z & $N_{tot}$ & $N_{SSP}$ & Mass(bf) & $\chi^2_{opt}$ & $\Delta M_{1\sigma}$ & $\Delta M_{2\sigma}$ & $\Delta M_{SIRTF}$\\
\# & & mdls. & b.f. & $[10^{11}M_\odot]$ & b.f. & $[10^{11}M_\odot]$ & $[10^{11}M_\odot]$ & $[10^{11}M_\odot]$\\
\hline
S14 & 0.41  & 420014 & 3 & 0.32 & 4.15 & 0.21$-$0.35  & 0.21$-$0.38  & \\
S15 & (0.55)& 451212 & 2 & 3.10 & 13.55& 2.97$-$3.29  & 2.61$-$3.49  & \\
S16 & 0.62  & 430819 & 4 & 0.29 & 4.73 & 0.23$-$0.28  & 0.22$-$0.31  & \\
S18 & (0.55)& 202806 & 4 & 5.20 & 14.37& 5.13$-$5.47  & 4.62$-$5.67  & \\
S20 & 0.39  & 416408 & 4 & 0.80 & 3.65 & 0.57$-$1.20  & 0.57$-$1.26  & \\
S23 & 0.46  & 409213 & 4 & 0.97 & 8.16 & 0.45$-$1.16  & 0.41$-$1.64  & 0.60$-$1.20\\
S25 & 0.58  & 430807 & 2 & 0.80 & 2.70 & 0.54$-$1.26  & 0.44$-$1.47  & 0.40$-$1.34\\
S27 & 0.58  & 507625 & 4 & 4.70 & 4.56 & 4.44$-$5.62  & 3.71$-$5.68  & 4.17$-$5.54\\
S28 & 0.58  & 476432 & 4 & 0.40 & 4.45 & 0.29$-$0.55  & 0.27$-$0.69  & \\
S30 & (0.40)& 206401 & 3 & 0.01 & 3.88 & 0.005$-$0.007& 0.005$-$0.025& \\
S36 & (0.65)& 198044 & 4 & 0.40 & 0.93 & 0.23$-$0.60  & 0.21$-$0.65  & \\
S40 & 1.27  & 385289 & 4 & 1.20 & 8.71 & 1.43$-$2.01  & 0.89$-$3.05  & 0.80$-$2.00\\
S41 & (0.30)& 208838 & 3 & 0.055& 6.43 & 0.051$-$0.059& 0.045$-$0.069& \\
S43 & 0.95  & 409212 & 4 & 0.50 & 0.77 & 0.25$-$0.87  & 0.22$-$0.94  & 0.21$-$0.77\\
S45 & (0.65)& 526817 & 4 & 0.80 & 5.67 & 0.53$-$1.69  & 0.24$-$1.71  & 0.30$-$1.25\\
S48 & (0.30)& 210015 & 3 & 0.10 & 4.08 & 0.14$-$0.21  & 0.058$-$0.21 & \\
S52 & (0.60)& 278402 & 4 & 0.13 & 3.81 & 0.13$-$0.17  & 0.12$-$0.17  & \\
S53 & 0.58  & 506477 & 3 & 1.20 & 3.34 & 0.99$-$1.38  & 0.99$-$1.52  & 1.20$-$1.38\\
S55 & 0.76  & 477664 & 5 & 1.30 & 1.16 & 0.87$-$1.69  & 0.55$-$1.69  & 1.15$-$1.90\\
S60 & 1.23  & 492090 & 4 & 2.50 & 14.38& 1.91$-$2.89  & 1.75$-$6.23  & 2.00$-$3.20\\
S62 & 0.73  & 462014 & 4 & 0.66 & 3.04 & 0.69$-$0.88  & 0.66$-$0.95  & \\
S67 & (1.00)& 424812 & 4 & 0.34 & 12.12& 0.21$-$0.87  & 0.21$-$0.91  & \\
S71 & (0.45)& 230406 & 4 & 0.04 & 5.02 & 0.016$-$0.044& 0.013$-$0.067 & \\
S72 & 0.55  & 421227 & 3 & 1.40 & 12.39& 1.14$-$1.30  & 1.12$-$2.01  & \\
S73 & 0.17  & 441608 & 4 & 1.50 & 10.00& 0.95$-$1.01  & 0.84$-$1.94  & 0.96$-$1.82\\
S75 & (0.45)& 222010 & 4 & 1.30 & 4.19 & 1.16$-$1.36  & 0.83$-$1.40  & \\
S77 & (0.40)& 210033 & 4 & 0.80 & 4.73 & 0.91$-$1.10  & 0.47$-$1.16  & 0.45$-$1.14\\
S79 & 0.74  & 499221 & 3 & 0.60 & 1.60 & 0.65$-$1.18  & 0.30$-$1.21 & \\
S82 & 0.69  & 438044 & 4 & 0.50 & 9.61 & 0.49$-$0.62  & 0.37$-$0.83  & 0.51$-$0.69\\
S85 & (0.40)& 414053 & 3 & 5.70 & 25.77& 5.51$-$5.79  & 4.78$-$6.19  & \\
\hline			    	  		      
\end{tabular}		    	  		      
\caption{Results of the stellar--mass estimate based of SEDs fitting. For each
ISO/LW3 HDFS source we report redshift (Franceschini et al. 2003; photometric if within 
parenthesis, spectroscopic otherwise), total number of models produced
by the spectrophotometric synthesis code (without the SIRTF canstraints),
effective number of SSPs contributing to best fit (for the optical+LW3 data, but
without the SIRTF constraints), optical $\chi^2$ and mass of the best fit
solution. The last three columns report three different estimates of   
the mass uncertainties due to degeneracies: $\Delta$M at 68.3\% and 95.5\% 
confidence levels (corresponding to 1$\sigma$ and 2$\sigma$) and 
2$\sigma$ mass range obtained simulating SIRTF/IRAC observations (only
for some randomly selected sources).}
\label{tab:results}
\end{table*}

\begin{table*}[!ht]
\centering
\begin{tabular}{l c c c c c c c c c c}
\hline
\hline
Obj & z & Color &$N_{tot}$ & $t_F$ & E(B--V) & $n_1$ & $\tau_1$ & $\chi^2$ & Mass(bf) & $\Delta M$ \\
\# & & Flag & mdls. & b.f. [Gyrs] & b.f. & b.f. & b.f. & b.f & $[10^{11}M_\odot]$ & $[10^{11}M_\odot]$ \\
\hline
E1  & 0.512   & b & 18001 & 11.91& 0.05 & 2.69  & 0.16 & 8.67 & 0.18 & 0.08$-$0.18\\
E2  & (1.3)   & b & 16848 & 11.66& 0.02 & 3.01  & 0.16 & 11.95& 0.83 & 0.47$-$1.07\\
E3  & (1.3)   & b & 23427 & 3.44 & 0.05 & 4.12  & 0.08 & 7.95 & 0.85 & 0.70$-$1.82\\
E4  & 0.414   & b & 11521 & 2.68 & 0.00 & 0.002 & 0.52 & 4.29 & 0.16 & 0.10$-$0.23\\
E5  & 0.673   & r & 129606& 3.19 & 0.05 & 0.001 & 0.15 & 1.89 & 0.73 & 0.66$-$1.12\\
E6  & (0.5)   & r & 111765& 9.20 & 0.06 & 0.002 & 0.15 & 6.31 & 0.90 & 0.52$-$1.16\\
E7  & 0.517   & b & 11287 & 4.63 & 0.10 & 5.17  & 0.11 & 6.71 & 0.10 & 0.08$-$0.14\\
E8  & 0.579   & r & 131529& 6.68 & 0.05 & 2.85  & 0.09 & 5.31 & 0.52 & 0.42$-$0.83\\
E9  & 0.515   & b & 15841 & 2.00 & 0.00 & 5.25  & 0.10 & 6.68 & 0.05 & 0.045$-$0.059\\
E11 & 0.364   & r & 121526& 7.17 & 0.04 & 2.35  & 0.06 & 9.92 & 0.10 & 0.07$-$0.16\\
E14 & (1.15)   & b & 11053 & 11.44& 0.00 & 1.97  & 0.25 & 5.52 & 0.75 & 0.43$-$0.81\\
E16 & (1.0)   & b & 11283 & 11.46& 0.05 & 0.04  & 0.26 & 8.36 & 3.42 & 2.29$-$3.99\\
E17 & (1.25)  & r & 136281& 11.71& 0.04 & 0.03  & 0.19 & 7.31 & 3.89 & 2.82$-$4.08\\
E18 & (0.9)   & r & 112240& 11.46& 0.04 & 0.02  & 0.19 & 4.58 & 0.73 & 0.38$-$0.89\\
E19 & (0.15)  & r & 121521& 11.96& 0.05 & 3.98  & 0.01 & 9.81 & 0.08 & 0.06$-$0.09\\
E20 & (0.8)   & b & 16087 & 2.00 & 0.00 & 0.001 & 0.34 & 9.76 & 0.12 & 0.08$-$0.18\\
E21 & 0.577   & r & 127454& 11.28& 0.05 & 0.04  & 0.29 & 10.00& 0.98 & 0.61$-$1.16\\
E24 & (0.4)   & r & 134001& 11.28& 0.05 & 3.35  & 0.08 & 13.54& 0.55 & 0.33$-$0.72\\
E25 & 0.511   & b & 13272 & 11.97& 0.04 & 1.49  & 0.21 & 7.02 & 0.72 & 0.44$-$0.93\\
E26 & (1.5)   & b & 11729 & 5.52 & 0.00 & 3.69  & 0.09 & 22.34& 2.80 & 1.89$-$4.12\\
E27 & (0.85)  & b & 11258 & 11.72& 0.02 & 3.42  & 0.11 & 16.02& 0.92 & 0.52$-$1.09 \\
E28 & 0.564   & b & 11281 & 11.91& 0.02 & 0.64  & 1.70 & 18.45& 0.13 & 0.086$-$0.14\\
E29 & (0.5)   & r & 112123& 14.35& 0.05 & 2.76  & 0.01 & 4.79 & 0.18 & 0.078$-$0.32\\
\hline
\end{tabular}
\caption{Results of the stellar--mass estimate for the ellipticals sample, 
based of SEDs fitting with solar metallicity SSPs. For each elliptical 
source in the HDFS (Rodighiero et al., 2001), spectroscopic (Vanzella et al. 2003, 
Sawicky \& Mallen-Ornelas 2003) or photometric 
(if within parenthesis, Rodighiero) redshifts, total number 
of models produced by the spectrophotometric synthesis code, best fit parameters
and $\chi^2$ for the analytic SFH method (see text for details),
mass of the best fit solution and 2$\sigma$ (95.5\% confidence) mass range due 
to degeneracies are reported. Third column reports a flag identifying red and
blue sources. The SEDs of the red sources have been fitted with three
different methods: {\em a)} analytic SFH, {\em b)} two single stellar populations 
(10$^9$ and 1.2 10$^{10}$ yrs old) with solar metallicity, {\em c)} two old
populations (1.2 10$^{10}$ yrs) with different metallicities (Z=0.02, 0.008) ---
see also text and figure \ref{fig:e_5}.
The mass ranges reported for these objects include the results of all three
fitting tecniques; on blue sources only the analytic SFH leads to good results.}
\label{tab:risultati_ellittiche}
\end{table*}

\begin{table*}[!ht]
\centering
\begin{tabular}{l c c c c c c c c c}
\hline
\hline
Obj & ID Pap. & ID Will. & z & $N_{tot}$ & $N_{SSP}$ & $\chi^2_{opt}$ & Mass(bf) & $\Delta M$ & $\Delta M_{SIRTF}$\\
\# & (1) & (2) & (3) & mdls. & b.f. & b.f. & $[10^{11}M_\odot]$ & $[10^{11}M_\odot]$ & $[10^{11}M_\odot]$ \\
\hline
Ly1 & 110 & 2-449.0 & 2.005 & 230403 & 3 & 5.13 & 0.21 & 0.16$-$2.47 & 0.27$-$0.55 \\
Ly2 & 503 & 2-903.0 & 2.233 & 195602 & 3 & 4.24 & 0.13 & 0.04$-$0.24 & 0.09$-$0.19 \\  
Ly3 & 67  & 2-82.1  & 2.267 & 195587 & 3 & 1.47 & 0.04 & 0.03$-$0.84 & 0.04$-$0.09 \\
Ly4 & 804 & 4-639.0 & 2.591 & 212778 & 2 & 10.99& 0.09 & 0.03$-$0.28 & 0.03$-$0.09 \\
Ly5 & 1352& 4-497.0 & 2.800 & 199203 & 3 & 5.87 & 0.02 & 0.02$-$0.29 & 0.02$-$0.06 \\
Ly6 & 1541& 4-363.0 & 2.980 & 217529 & 3 & 9.02 & 0.18 & 0.05$-$0.35 & 0.06$-$0.20 \\
Ly7 & 661 & 2-643.0 & 2.991 & 264435 & 2 & 7.54 & 0.04 & 0.02$-$0.13 & 0.03$-$0.06 \\
Ly8 & 273 & 2-76.11 & 3.160 & 278399 & 2 & 2.29 & 0.56 & 0.27$-$1.35 & 0.38$-$0.81 \\
\hline			    	  		      
\multicolumn{10}{l}{(1): ID number in Papovich et al. (2001);}\\
\multicolumn{10}{l}{(2): ID number in Williams et al. (1996);}\\
\multicolumn{10}{l}{(3): Spectroscopic redshifts by Steidel et al (1996) and Lowental et al. (1997).}\\
\end{tabular}		    	  		      
\caption{Results of the spectral synthesis analysis of the lyman-break galaxies.
Objects are identified by numbers adopted in this work, by Papovich and by
Williams (first three columns). For each source the table reports redshift,
total number of models analyzed, effective number of SSPs involved in the best
fit, stellar mass and $\chi^2$ of best fit and 2$\sigma$ mass range due to
degeneracies. Last column summarizes the results of our simulated SIRTF/IRAC
observations of ly-break sources: IRAC will sample the restframe near-IR light
at $z\sim 2-3$, therefore significantly reducing the uncertainty on the
stellar mass estimate (see text for details).}
\label{tab:results_ly}
\end{table*}

%%%%%%%%%%%%%%%%%%%%%%%%%%%%%%%%%%%%%%%%%%%%%%%%%%%%%%%%%%%%%%%%%%%%%%%%%% 
% bibliography 

% tabelle con masse e ranges
%\include{tab_masses}

%\include{tab_masses_ellittiche_ly}
%\include{tab_masses_ellittiche}
%\include{tab_masses_ly}

%\include{tab_indet}

%%%%%%%%%%%%%%%%%%%%%%%%%%%%%%%%%%%%%%%%%%%%%%%%%%%%%%%%%%%%%%%%%%%%%%%%%%
\end{document}